\documentclass[conference]{IEEEtran}
\IEEEoverridecommandlockouts
\usepackage{cite}
\usepackage{amsmath,amssymb,amsfonts}
\usepackage{algorithmic}
\usepackage{graphicx}
\usepackage{textcomp}
\usepackage{xcolor}
\def\BibTeX{{\rm B\kern-.05em{\sc i\kern-.025em b}\kern-.08em
    T\kern-.1667em\lower.7ex\hbox{E}\kern-.125emX}}
\begin{document}

\title{Energy Efficiency Optimization of Generalized Spatial Modulation with Sub-Connected Hybrid Precoding}

\author{\IEEEauthorblockN{Kai Chen\IEEEauthorrefmark{2}, Jing Yang\IEEEauthorrefmark{2}, Xiaohu Ge\IEEEauthorrefmark{2}\IEEEauthorrefmark{1}, Yonghui Li\IEEEauthorrefmark{3}, Lin Tian\IEEEauthorrefmark{4},
Jinglin Shi\IEEEauthorrefmark{4}\IEEEauthorrefmark{5}}
\IEEEauthorblockA{\IEEEauthorrefmark{2}School of Electronic Information and Communications, Huazhong University of Science and Technology, Wuhan, Hubei,\\ China}
\IEEEauthorblockA{\IEEEauthorrefmark{3}School of Electrical and Information Engineering, University of Sydney, Sydney, Australia}
\IEEEauthorblockA{\IEEEauthorrefmark{4}Beijing Key Laboratory of Mobile Computing and Pervasive Devices, Institute of Computing Technology, Chinese\\
 Academy of Sciences, China}
\IEEEauthorblockA{\IEEEauthorrefmark{5}\textit{University of Chinese Academy of Sciences}, China\\}
e-mail: xhge@mail.hust.edu.cn
}

\maketitle

\begin{abstract}
Energy efficiency (EE) optimization of millimeter wave (mm-Wave) massive multiple-input multiple-output (MIMO) systems is emerging as an important challenge for the fifth generation (5G) mobile communication systems. However, the power of radio frequency (RF) chains increases sharply due to the high carrier frequency in mm-Wave massive MIMO systems. To overcome this issue, a new energy efficiency optimization solution is proposed based on the structure of the generalized spatial modulation (GSM) and sub-connected hybrid precoding (HP). Moreover, the computation power of mm-Wave massive MIMO systems is considered for optimizing the EE. Simulation results indicate that the EE of the GSM-HP scheme outperforms the full digital precoding (FDP) scheme in the mm-Wave massive MIMO scene, and 88\% computation power can be saved by the proposed GSM-HP scheme.
\end{abstract}


\section{Introduction}
Millimeter wave (mm-Wave) communication and massive multiple-input multiple-output (MIMO) are two key technologies in the fifth generation (5G) mobile communication systems \cite{17Ge}. Despite the fact that the mm-Wave technology can enable high-rate communication, fast attenuation and short communication distance will limit the performance of mm-Wave communication systems \cite{1Khan}. Fortunately, massive MIMO technology can provide high beamforming gain, which can concentrate the transmission energy in a certain direction and overcome the attenuation problem of mm-Wave transmission \cite{2Taeyoung}\cite{24Ge}.

Although the 5G mm-Wave massive MIMO systems can increase the spectrum efficiency, the large number of RF chains in the massive MIMO systems will consume huge energy and restrict the improvement of the EE \cite{3Gao}\cite{4Zi}. To reduce the energy consumption of RF chains, hybrid precoding (HP) technology is proposed to lessen the number of RF chains in the massive MIMO systems \cite{5Zhang}\cite{6Alkhateeb}.There are two typical hybrid precoding structures currently, i.e. the sub-connected structure and the fully-connected structure. The sub-connected array adopts the structure that each RF chain is merely connected to part of the antennas through a phase shifter. In this way, system complexity is reduced at the cost of the antenna gain, which is more valuable for improving the EE.

Generalized spatial modulation (GSM) is another technology that has shown promise in reducing the number of RF chains. By utilizing the additional spatial dimension, GSM can improve the EE of massive MIMO system while guaranteeing the spectral efficiency \cite{7Renzo}. The input data streams in GSM are divided into two parts, one part is the amplitude-phase modulation (APM) domain data streams for the traditional N-order symbol modulation, and the other part is the space-domain data stream for the antenna selection \cite{8Younis}\cite{9Lakshmi}. GSM takes the index of activated antennas as spatial symbols that exploit degree of spatial freedom without introducing any RF chains. Therefore, combining GSM with mm-Wave massive MIMO system can effectively reduce the power consumption and hardware complexity.

Most of the studies regarding the EE of cellular networks ignore the computation power of the base station or simply set it to a constant value \cite{20Ge}\cite{21Humar}\cite{22Ge}\cite{23Ge}. But \cite{10Ge} points out that the computation power of the massive MIMO system will consume more than 50\% energy at the base station, and the optimization of computation power plays a major role in the improvement of the EE for 5G cellular networks. In this paper, we study the EE of the system combining GSM with HP technology in the mm-Wave scene, and the computation power is considered in the analysis. The EE is analyzed with respect to the number of active users, RF chains, and antennas per group via simulations. Note that the GSM-HP system was first proposed in \cite{11He}, in which spectral efficiency was studied. To the best of our knowledge, this paper is the first to research the EE of the GSM-HP system in the mm-Wave massive MIMO scene.

The remainder of this paper is organized as follows. The system model of the GSM with sub-connected HP scheme is described in Section II. Section III analyzes the EE of the system. In Section IV, simulation results are provided, and Section V concludes this paper.

\section{System Model}
\begin{figure}[htbp]
\centerline{\includegraphics[width=0.5\textwidth]{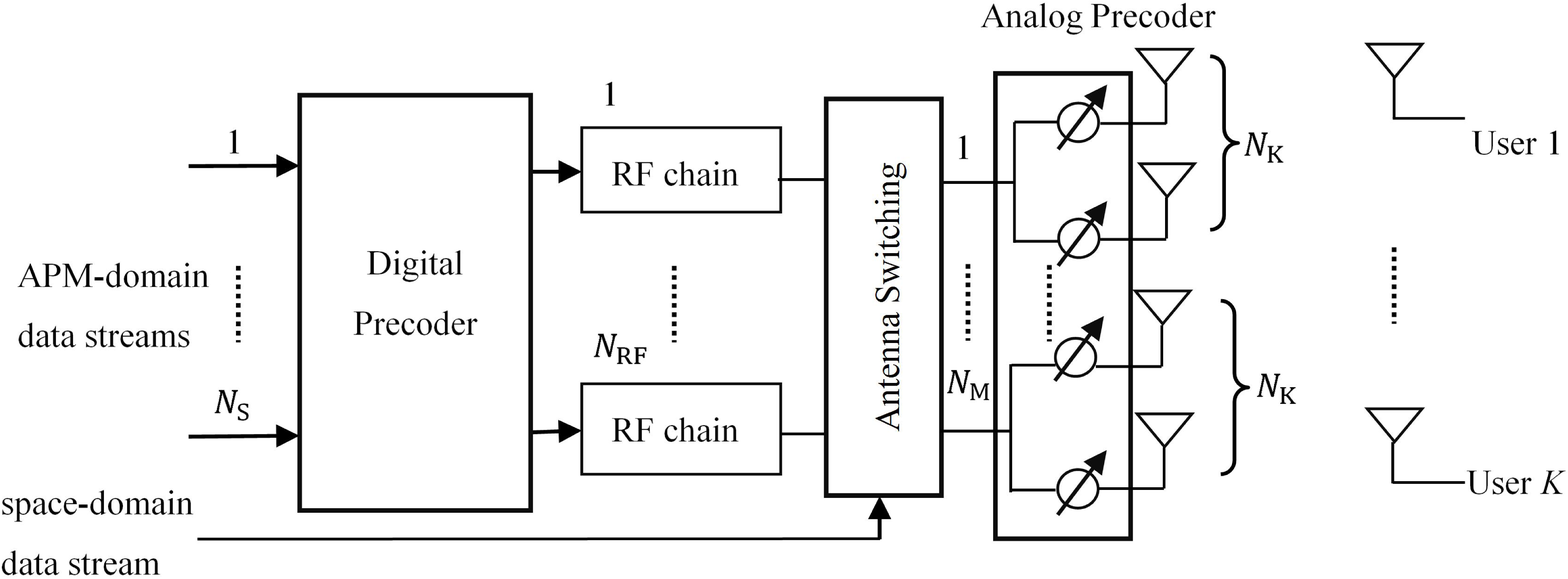}}
\caption{System model of the GSM with sub-array HP scheme.}
\label{fig1}
\end{figure}

In this paper, we analyze the EE of the GSM with sub-connected HP scheme. The block diagram is given in Fig.~\ref{fig1}, where a mm-Wave massive MIMO system with ${{N}_\text{T}}$ transmit antennas and $K$ single-antenna users is considered.

The input of the system includes $\text{1}$ space-domain data stream and ${{N}_{\text{S}}}=K$ APM-domain data streams. The ${{N}_{\text{S}}}$ APM-domain data streams are then transmitted to the baseband precoder to generate ${{N}_\text{RF}}$ RF-domain symbols. Since the sub-connected hybrid precoding structure is adopted, we divide the ${{N}_\text{T}}$ transmit antennas into ${{N}_\text{M}}$ antenna groups, and each group consists of ${{N}_{\text{K}}}$ antennas, which satisfies ${{N}_\text{T}}={{N}_\text{M}}{{N}_{\text{K}}}$. According to the principle of GSM, it is required that ${{N}_\text{S}}\le {{N}_\text{RF}}<{{N}_\text{M}}$. Therefore, the space-domain date stream can control the switches to distribute the output signals of RF chains to ${{N}_\text{RF}}$ out of the ${{N}_\text{M}}$ antenna groups, while the remaining $\left( {{N}_\text{M}}-{{N}_\text{RF}} \right)$ antenna groups will not be activated during the signal transmission period.

The number of available spatial modulation schemes is symbolized by ${M}\ge {1}$. Each spatial modulation scheme is determined by the space-domain data stream, ${M}$ can thus be expressed as:

\begin{equation}
{M}={{{2}}^{\left\lfloor {\text{log}_{{2}}}\left( \begin{matrix}
   {{N}_\text{M}}  \\
   {{N}_\text{RF}}  \\
\end{matrix} \right) \right\rfloor }},
\label{eq}
\end{equation}
where $\left( _{\centerdot}^{\centerdot} \right)$ denotes the binomial coefficient and $\left\lfloor \centerdot  \right\rfloor $ denotes the floor operation. We let ${\mathbf{C}_\text{m}}\in {{\mathbb{C}}^{{{N}_{\text{T}}}\times {{N}_{\text{RF}}}}}$ represent the $m$-th spatial modulation matrix with ${1 \le m\le {M}}$. Due to the space limitation, the explanation for the spatial modulation matrix is omitted here, and more technical details can be found in \cite{11He}. Finally, the received signal vector $\mathbf{y}\in {{\mathbb{C}}^{K\times {1}}}$ is given as follows when the $m$-th spatial modulation scheme is selected:

\begin{equation}
{\mathbf{y}}={\mathbf{H}^{\text{H}}}\mathbf{A}{\mathbf{C}_\text{m}}{\mathbf{D}_\text{m}}\mathbf{x}+\mathbf{n},
\label{eq}
\end{equation}
where $\mathbf{A}\in {{\mathbb{C}}^{{{N}_{\text{T}}}\times {{N}_{\text{T}}}}}$ represents the RF precoding matrix, and ${\mathbf{D}_\text{m}}\in {{\mathbb{C}}^{{{N}_{\text{RF}}}\times{{N}_{\text{S}}}}}$ represents the baseband precoding matrix when the $m$-th spatial modulation matrix is selected. Limited by the total transmitting power ${{P}_\text{max}}$ of the base station, the precoding matrix is required to satisfy:

\begin{equation}
\left\| \mathbf{A}{\mathbf{C}_\text{m}}{\mathbf{D}_\text{m}} \right\|_{F}^{2}={{P}_{\text{max}}},
\label{eq}
\end{equation}
where ${{\left\| \centerdot  \right\|}_{F}}$ denotes the Frobenius norm of the matrix.

We let $\mathbf{x}={{\left[ {{x}_1},\ldots ,{{{x}}_\text{k}},\ldots ,{{{x}}_{{{N}_{\text{S}}}}} \right]}^{{T}}}$ symbolize the ${{N}_{\text{S}}}$ APM-domain data streams, where ${{{x}}_\text{k}}$ obeys a complex Gaussian distribution with 0-mean and 1-variance, and $\mathbf{n}={{\left[ {{{n}}_{{1}}},\ldots ,{{{n}}_\text{k}},\ldots ,{{{n}}_\text{K}} \right]}^{{T}}}$ symbolize the additive white Gaussian noise (AWGN), where ${{n}_\text{k}}$ obeys a complex Gaussian distribution with 0-mean and $\sigma _{\text{N}}^{2}$-variance. ${{\left[ \centerdot  \right]}^{{T}}}$ denotes the transposition.

${\mathbf{H}^{\text{H}}}\in {{\mathbb{C}}^{K\times {{N}_{\text{T}}}}}$ is the mm-Wave massive MIMO channel matrix, and $\mathbf{h}_\text{k}^\text{H}\in {{\mathbb{C}}^{{1}\times {{N}_{\text{T}}}}}$ represents the channel matrix of the $k$-th user. Obviously, ${\mathbf{H}^{\text{H}}}={{\left[ {\mathbf{h}_{1}},\ldots ,{\mathbf{h}_\text{k}},\ldots ,{\mathbf{h}_\text{K}} \right]}^{\text{H}}}$, where ${{\left[ \centerdot  \right]}^{\text{H}}}$ denotes the conjugate transposition. Due to the high-path-loss propagation characteristic in free space, the spatial selection and scattering of mm-Wave are limited. Meanwhile, the large-scale antenna array structure of the mm-Wave transceiver leads to high correlation of the antenna, and the number of propagation paths of mm-Wave is much smaller than that of transmission antennas. Therefore, it is not accurate to model the mm-Wave channel with the statistical fading distribution that used in conventional MIMO analysis in a sparse scattering environment. In this paper, we adopt the geometric channel model with a finite scattering and multipath to characterize the mm-Wave MIMO channel \cite{12Ayach}. The channel matrix of the $k$-th user is given as:

\begin{equation}
{\mathbf{h}_\text{k}}=\sqrt{\frac{{{N}_{\text{T}}}{{\beta }_\text{k}}}{{{N}_{\text{ray}}}}}\sum\limits_{{i}=1}^{{{N}_{\text{ray}}}}{{{\rho }_\text{ki}}\mathbf{u}\left( {{\psi }_\text{i}},{{\vartheta }_\text{i}} \right)},
\label{eq}
\end{equation}
where ${{N}_{\text{ray}}}$ represents the number of multipath between users and the base station, and ${{N}_{\text{T}}}$ represents the number of transmission antennas. ${{{\beta }_\text{k}}=\zeta /l_\text{k}^{\gamma }}$ is the large-scale fading coefficient between the base station and the $k$-th user, where $\zeta $ obeys a lognormal distribution with 0-mean and 9.2 dB-variance and ${l}_{k}$ represents the distance between the $k$-th user and the base station. $\gamma $ is the path loss factor and is set as 4.6 in this paper \cite{18Ge}. ${{\rho }_\text{ki}}$ is the complex gain of the $k$-th user on the $i$-th multipath, which is considered as the small-scale fading coefficient. Furthermore, ${{\rho }_\text{ki}}$ is an independent and identically-distributed random variable for each $k(1,\ldots ,K)$ and $i(1,\ldots ,{{N}_\text{ray}})$. ${{\psi }_\text{i}}$ and ${{\vartheta }_\text{i}}$ represent the azimuth and elevation angle of the $i$-th multipath between users and the base station from the antenna array of the base station, respectively. Compared with other antenna structures, the two-dimensional planar antenna array is smaller in size and less complex to make the most of the angle information of the signal, which is more suitable for a mm-Wave massive MIMO system. Therefore, we employ the planar antenna array structure in this paper. The array response ${\mathbf{u}\left( {{\psi }_\text{i}},{{\vartheta }_\text{i}} \right)}$ corresponding to azimuth angle ${{\psi }_\text{i}}$ and elevation angle ${{\vartheta }_\text{i}}$ is formulated as follows:

\begin{equation}
\begin{gathered}
{\mathbf{u}}( {{\psi _\text{i}},{\vartheta _\text{i}}} ) = \frac{1}{{\sqrt {{N_{\text{T}}}} }}[1, \ldots ,\exp j\frac{{2\pi }}{\lambda }d( l\sin ( {{\psi _\text{i}}} )\sin( {{\vartheta _\text{i}}} ) \hfill \\
\;\;\;\;\;\;\;\;\;\;\;\;\;\;+ r\cos( {{\vartheta _\text{i}}}) ),...,\exp j\frac{{2\pi }}{\lambda }d(( {L - 1} )\sin ( {{\psi _\text{i}}} )\sin( {{\vartheta _\text{i}}} ) \hfill \\
\;\;\;\;\;\;\;\;\;\;\;\;\;\;+ ( {R - 1} )\cos( {{\vartheta _\text{i}}} ))]^\text{T},
\label{eq}
\end{gathered}
\end{equation}
where $\lambda $ represents the carrier wavelength and $d=\lambda/2$ represents the inter-element spacing. Besides, $0\le l\le \left( L-1 \right)$ and $0\le r\le \left( R-1 \right)$ are the row and column of antenna array, and the size of the antenna array is ${{N}_{\text{T}}}=LR$.

\section{Energy Efficiency Optimization}

In this paper, the energy efficiency ${{\eta }_{\text{EE}}}$ of mm-Wave massive MIMO systems is expressed as \cite{19Xiang}

\begin{equation}
{{\eta }_{\text{EE}}}=\frac{{{R}_{\text{total}}}}{{{P}_{\text{total}}}},
\label{eq}
\end{equation}
where ${{R}_{\text{total}}}$ is the wireless channel capacity of mm-Wave massive MIMO systems, and ${{P}_{\text{total}}}$ is the total power of mm-Wave massive MIMO systems.

\subsection{Wireless Channel Capacity}
According to the system model in Section II, the received signal of the  $k$-th user is given as

\begin{equation}
{\mathbf{y}_{\text{k}}}=\mathbf{h}_{\text{k}}^\text{H}\mathbf{A}{\mathbf{C}_{\text{m}}}{\mathbf{D}_{\text{m}}}\mathbf{x}+{{n}_{\text{k}}}.
\label{eq}
\end{equation}

The spectral efficiency ${{R}_{\text{k}}}$ of single-user GSM is calculated by the mutual information between $\mathbf{x}$, $m$ and ${{\mathbf{y}}_{\text{k}}}$, i.e.
\begin{equation}
{{{R}_{\text{k}}}=I\left( {\mathbf{y}_{\text{k}}};\mathbf{x},m \right)}.
\label{eq}
\end{equation}

Since $m$ is a discrete channel input, the mutual information above cannot be expressed in a closed form, which brings great inconvenience to the performance analysis. In \cite{11He}, an approximate closed-form expression of spectral efficiency is provided:

\begin{equation}
{{{R}_{\text{k}}}=\text{log}_{2}\frac{{M}}{{2}\sigma_{\text{N}}^{{2}}}-\frac{1}{{M}}\sum\limits_{{n}=1}^{M}{\text{log}_2\left( \sum\limits_{{t}=1}^{M}{{{\left| {{\sum }_{\text{k,n}}}+{{\sum }_{\text{k,t}}} \right|}^{-1}}} \right)}},
\label{eq}
\end{equation}
where ${{\sum }_{\text{k,n}}}$ symbolizes the conditional covariance matrix of ${\mathbf{y}_{\text{k}}}$ when the $n$-th spatial modulation matrix is chosen. ${{\sum }_{\text{k,n}}}$ is given as follows:

\begin{equation}
{{{\sum }_{\text{k,n}}}\triangleq \sigma _{\text{N}}^{2}+\mathbf{h}_{\text{k}}^\text{H}\mathbf{A}{\mathbf{C}_{\text{n}}}{\mathbf{d}_{\text{n,k}}}\mathbf{d}_{\text{n,k}}^{\text{H}}\mathbf{C}_{\text{n}}^{\text{H}}{\mathbf{A}^{\text{H}}}{\mathbf{h}_{\text{k}}}}.
\label{eq}
\end{equation}

Based on the spectrum efficiency of the  $k$-th user, the total channel capacity can be expressed as

\begin{equation}
{{R}_{\text{total}}}=B\sum\limits_{{k}=1}^{K}{{{R}_{\text{k}}}},
\label{eq}
\end{equation}
where $B$ represents the bandwidth.

A reasonable hybrid precoding algorithm can achieve the same system performance as the optimal full-digital precoding (FDP)\cite{13Guo}.Therefore, to simplify the analysis, we assume the performance of HP consistent with full-digital zero-forcing precoding when calculating the channel capacity.

\subsection{Power Consumption}

As mentioned in Section I, the computation power cannot be ignored or set to a constant in the mm-Wave massive MIMO scene. Similar to \cite{14Bjornson}, we divide the total power of the base station into three parts, including transmission power, computation power and the fixed power:
\begin{equation}
{{{P}_{\text{total}}}={{P}_{\text{transmission}}}+{{P}_{\text{computation}}}+{{P}_{\text{fix}}}}.
\label{eq}
\end{equation}

The detailed power consumption is modeled below based on the system introduced in Section II.

\subsubsection{Transmission Power}
Transmission power consists of power consumed by amplifiers, radio frequency and switches.

The amplifier power ${{P}_{\text{PA}}}$ is expressed as
\begin{equation}
{{{P}_{\text{PA}}}=\frac{1}{\alpha }\left\| \mathbf{A}{\mathbf{C}_\text{m}}{\mathbf{D}_\text{m}} \right\|_{F}^{2}=\frac{{{P}_{\text{max}}}}{\alpha }},
\label{eq}
\end{equation}
where $\alpha $ stands for the efficiency of the amplifier.

The RF power includes the power of RF chains and phase shifters in the RF precoding. Considering that there are only ${{N}_{\text{RF}}}$ out of ${{N}_{\text{M}}}$ antenna groups chosen to work in the GSM system, hence the number of working antennas is ${{N}_{\text{RF}}}{{N}_{\text{K}}}$, which is equal to that of phase shifters. With ${{P}_{\text{RF}\_\text{per}\_\text{chain}}}$ symbolizing the power of each RF chain and ${{P}_{\text{per}\_\text{shifter}}}$ symbolizing the power of each phase shifter, the RF power is formulated as:

\begin{equation}
{{{P}_{\text{RF}}}={{N}_{\text{RF}}}{{P}_{\text{RF}\_\text{per}\_\text{chain}}}+{{N}_{\text{RF}}}{{N}_{\text{K}}}{{P}_{\text{per }\!\!\_\!\!\text{ shifter}}}}.
\label{eq}
\end{equation}

Switches are required to select the antenna groups in the spatial modulation structure, the switch power can thus be expressed as
\begin{equation}
{{{P}_{\text{switch}}}={{N}_{\text{RF}}}{{P}_{\text{per}\_\text{switch}}}}.
\label{eq}
\end{equation}

According to (13) (14) (15), we then obtain the transmission power:

\begin{equation}
\begin{gathered}
{{P}_{\text{transmission}}}=\frac{{{P}_{\text{max}}}}{\alpha }+{{N}_{\text{RF}}}{{P}_{\text{RF}\text{ }\!\!\_\!\!\text{ }\text{per}\text{ }\!\!\_\!\!\text{ }\text{chain}}} \hfill \\
\;\;\;\;\;\;\;\;\;\;\;\;\;\;\;\;\;+{{N}_{\text{RF}}}{{N}_{\text{K}}}{{P}_{\text{per }\!\!\_\!\!\text{ shifter}}}+{{N}_{\text{RF}}}{{P}_{\text{per}\_\text{switch}}}.
\label{eq}
\end{gathered}
\end{equation}

\subsubsection{Computation Power}
Computation power is composed of all the power consumed by the base station for calculation, including channel estimation, channel coding, and linear processing.

The channel estimation is processed within a stable coherent block, therefore, the channel estimation power is expressed as the product of the number of coherent blocks unit time ${{\nu }_{\text{block}}}$ and the energy consumption per channel estimation $\kappa$, i.e.

\begin{equation}
{{{P}_{\text{CE}}}={{\nu }_{\text{block}}}\kappa},
\label{eq}
\end{equation}
where ${{\nu }_{\text{block}}}$ is calculated by the coherence time ${{T}_{\text{c}}}$ and the coherence bandwidth ${{B}_{\text{c}}}$, i.e.

\begin{equation}
{{{\nu }_{\text{block}}}=\frac{B}{{{B}_{\text{c}}}{{T}_{\text{c}}}}}.
\label{eq}
\end{equation}

$\kappa$ in (17) is given as
\begin{equation}
{{\kappa}=\frac{{{\gamma }_{\text{CE}}}}{{{L}_{\text{BS}}}}},
\label{eq}
\end{equation}
where ${{\gamma }_{\text{CE}}}$ represents the floating-point operations needed per channel estimation and ${{L}_{\text{BS}}}$ represents computation efficiency (in Gigaflops/Watt) of the base station. Assuming that the pilot-based channel estimation method is adopted, then ${{N}_{\text{T}}}$ pilot sequences will be received at the base station. The length of each pilot sequence is $\tau K$, where $\tau \ge 1$ denotes the factor that enables the pilots to be orthogonal. The base station estimates the channel according to the product of the pilot and pilot sequences with the length of $\tau K$ \cite{15Mohammed}, ${{\gamma }_{\text{CE}}}$ can thus be formulated as

\begin{equation}
{{{\gamma }_{\text{CE}}}=2\tau {{N}_\text{T}}{{K}^{2}}}.
\label{eq}
\end{equation}

According to (18) (19) (20), we obtain channel estimation power:
\begin{equation}
{{{P}_{\text{CE}}}=\frac{B}{{{B}_{\text{c}}}{{T}_{\text{c}}}}\frac{2\tau {{N}_{\text{T}}}{{K}^{2}}}{{{L}_{\text{BS}}}}}.
\label{eq}
\end{equation}

Channel coding power is proportional to the information rate, which is written as

\begin{equation}
{{{P}_{\text{CD}}}={{P}_{\text{COD}}}{{R}_{\text{total}}}={{P}_{\text{COD}}}B\sum\limits_{{k}=1}^{K}{{{R}_{\text{k}}}}},
\label{eq}
\end{equation}
where ${{P}_{\text{COD}}}$ represents the efficiency of channel coding (in Watt per bit/s).

In this paper, linear processing includes baseband precoding and solution of the precoding matrix. The power of baseband precoding is expressed as:

\begin{equation}
{{{P}_{\text{BB}}}=B\frac{\gamma }{{{L}_{\text{BS}}}}},
\label{eq}
\end{equation}
where $\gamma$ represents the floating-point operations per baseband precoding. The product of the baseband precoding matrix (${{N}_{\text{RF}}}\times {{N}_{\text{S}}}$) and the data stream matrix (${{N}_{\text{S}}}\times 1$) requires $2{{N}_{\text{RF}}}{{N}_{\text{S}}}$ floating-point operations \cite{16Hunger}. Considering that the baseband signal is in the complex domain, we modify the floating-point operations as

\begin{equation}
{{\gamma} =8{{N}_{\text{RF}}}{{N}_{\text{S}}}}.
\label{eq}
\end{equation}

According to (23) (24), the power of baseband precoding is given by

\begin{equation}
{{{P}_{\text{BB}}}=B\frac{\text{8}{{N}_{\text{RF}}}{{N}_{\text{S}}}}{{{L}_{\text{BS}}}}}.
\label{eq}
\end{equation}

The solution of the precoding matrix is carried out evert time channel estimation is processed, the power consumed by the solution of the precoding matrix can thus be expressed as the product of the number of coherent blocks unit time ${{\nu }_{\text{block}}}$ and the energy consumption per solution operation ${{\kappa }_\text{precoding}}$:
\begin{equation}
{{{P}_{\text{LP }\!\!\_\!\!\text{ C}}}={{\nu }_{\text{block}}}{{\kappa }_\text{precoding}}},
\label{eq}
\end{equation}
where ${{\kappa }_\text{precoding}}$ can be expressed as the ratio of floating-point operations required to perform a solution of the precoding matrix to the computation efficiency of the base station, i.e.
\begin{equation}
{{{\kappa }_\text{precoding}}=\frac{{{\gamma }_{\text{precoding}}}}{{{L}_{\text{BS}}}}}.
\label{eq}
\end{equation}

The power consumed by the solution operation of the precoding matrix can then be concluded as:
\begin{equation}
{{{P}_{\text{LP }\!\!\_\!\!\text{ C}}}=\frac{B}{{{B}_{\text{c}}}{{T}_{\text{c}}}}\frac{{{\gamma }_{\text{precoding}}}}{{{L}_{\text{BS}}}}}.
\label{eq}
\end{equation}

According to (25) (28), we obtain the linear processing power:
\begin{equation}
{{{P}_{\text{LP}}}=B\frac{8{{N}_{\text{T}}}{{N}_{\text{S}}}}{{{L}_{\text{BS}}}}+\frac{B}{{{B}_{\text{c}}}{{T}_{\text{c}}}}\frac{{{\gamma }_{\text{precoding}}}}{{{L}_{\text{BS}}}}}.
\label{eq}
\end{equation}

Moreover, according to (21)(22)(29), the computation power can be expressed as:

\begin{equation}
\begin{gathered}
{{P}_{\text{computation}}}=\frac{B}{{{B}_{\text{c}}}{{T}_{\text{c}}}}\frac{2\tau {{N}_{\text{T}}}{{K}^{{2}}}}{{{L}_{\text{BS}}}}+{{P}_{\text{COD}}}B\sum\limits_{{k}={1}}^{K}{{{R}_{\text{k}}}} \hfill \\
\;\;\;\;\;\;\;\;\;\;\;\;\;\;\;\;+B\frac{{8}{{N}_{\text{T}}}{{N}_{\text{S}}}}{{{L}_{\text{BS}}}}+\frac{B}{{{B}_{\text{c}}}{{T}_{\text{c}}}}\frac{{{\gamma }_{\text{precoding}}}}{{{L}_{\text{BS}}}}.
\label{eq}
\end{gathered}
\end{equation}

\subsubsection{Fixed Power}
Other power, such as cooling, voltage conversion loss, etc., is set to the fixed power ${{P}_{\text{fix}}}$.

Finally, by submitting (16) and (30) into (12), the total power of the base station is given as follows:

\begin{equation}
\begin{gathered}
{{P}_{\text{total}}}=\frac{{{P}_{\text{max}}}}{\alpha }+{{N}_{\text{RF}}}{{P}_{\text{RF}\text{ }\!\!\_\!\!\text{ }\text{per}\text{ }\!\!\_\!\!\text{ }\text{chain}}}
+{{N}_{\text{RF}}}{{N}_{\text{K}}}{{P}_{\text{per }\!\!\_\!\!\text{ shifter}}}\hfill \\
\;\;\;\;\;\;\;\;+{{N}_{\text{RF}}}{{P}_{\text{per}\_\text{switch}}}
+\frac{B}{{{B}_{\text{c}}}{{T}_{\text{c}}}}\frac{{2}\tau {{N}_{\text{T}}}{{K}^{{2}}}}{{{L}_{\text{BS}}}}+{{P}_{\text{COD}}}{{R}_{\text{total}}} \hfill \\
\;\;\;\;\;\;\;\;+B\frac{{8}{{N}_{\text{RF}}}{{N}_{\text{S}}}}{{{L}_{\text{BS}}}}+\frac{B}{{{B}_{\text{c}}}{{T}_{\text{c}}}}\frac{{{\gamma }_{\text{precoding}}}}{{{L}_{\text{BS}}}}+{{P}_{\text{fix}}}.
\label{eq}
\end{gathered}
\end{equation}

\section{Simulation Results}
In this section, we present the simulation results characterizing the EE of the considered GSM-HP system compared with traditional full-digital zero-forcing precoding system. In particular, the computation power is taken into account when calculating EE. The simulation parameters are given in Table I.

\begin{table}[htbp]
\caption{Simulation Paraneters}
\begin{center}
\begin{tabular}{|c|c|}
\hline
\textbf{Parameters}& \textbf{Values} \\
\hline
Transmitting power ${{P}_{\text{max}}}$ & 39 dBm \\
\hline
Power spectral density (PSD) of noise  & -174 dBm/Hz \\
\hline
Efficiency of power amplifier $\alpha $ & 0.38 \\
\hline
Bandwidth of mm-Wave $B$ & 800 MHz \\
\hline
Power per RF chain ${{P}_{\text{RF}\_\text{per}\_\text{chain}}}$ & 45 mW \\
\hline
Carrier frequency ${{f}_{\text{c}}}$  & 28 GHz \\
\hline
Coherent time ${{T}_{\text{c}}}$  & 5 ms \\
\hline
Coherent bandwidth ${{B}_{\text{c}}}$  & 100 MHz \\
\hline
$\tau$  & 1 \\
\hline
Computation efficiency of base station ${{L}_{\text{BS}}}$  & 12.8 Gigaflops/Watt \\
\hline
Efficiency of channel coding ${{P}_{\text{COD}}}$ & $10^{-10} \text{ Watt per bit/s}$  \\
\hline
power per phase shifter ${{P}_{\text{per}\_\text{shifter}}}$  & 15 mW \\
\hline
power per switch ${{P}_{\text{per}\_\text{switch}}}$  &  5 mW \\
\hline
fixed power of base station ${{P}_{\text{fix}}}$ & 1 W \\
\hline
Number of multipath ${{N}_{\text{ray}}}$ & 20 \\
\hline
Path loss factor $\gamma $ & 4.6 \\
\hline
Azimuth angle ${{\psi }_\text{i}}$ and elevation angle ${{\vartheta }_{\text{i}}}$  & $\left[ 0,2\pi  \right]$ uniform distribution \\
\hline
\end{tabular}
\label{tab1}
\end{center}
\end{table}

\begin{figure}[htbp]
\centerline{\includegraphics[width=0.5\textwidth]{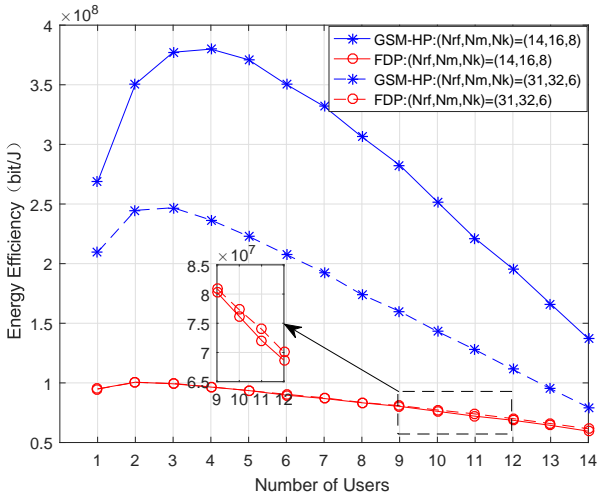}}
\caption{Energy efficiency VS. number of users of different schemes.}
\label{fig2}
\end{figure}

In Fig.~\ref{fig2}, we plot the EE of GSM-HP system and FDP system with respect to the number of users. It can be observed that the EE of FDP goes down as the number of users increases. Huge energy consumption, especially computation power is required by more users, while the FDP system cannot provide the corresponding capacity gain in the mm-Wave scene. The EE of GSM-HP is positively correlated with the number of users at the first beginning, and vanishes as more users are added, but is still superior to FDP. Spatial modulation technology utilizes spatial freedom to increase channel capacity without bringing in extra power consumption. Therefore, when there are not too many users, the capacity gain provided by spatial modulation can compensate for the decline in energy efficiency of the system. But the gain is finite, the EE decreases as users increases furthermore.

\begin{figure}[htbp]
\centerline{\includegraphics[width=0.5\textwidth]{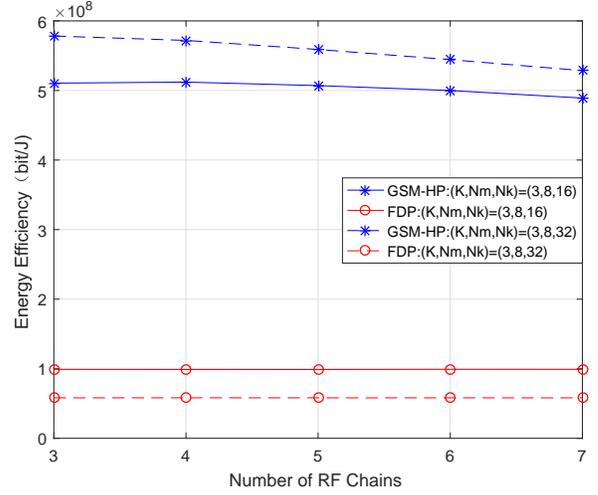}}
\caption{Energy efficiency VS. number of RF chains of different schemes.}
\label{fig3}
\end{figure}

The EE against the number of RF chains is investigated in Fig.~\ref{fig3}. In this part of simulation, RF chains of the FDP scheme are required to be equal to the number of antennas, therefore, the EE of FDP is a constant value. While the EE of GSM-HP declines as the number of RF chains increases, indicating that the increasing RF chains bring in more energy consumption.

\begin{figure}[htbp]
\centerline{\includegraphics[width=0.5\textwidth]{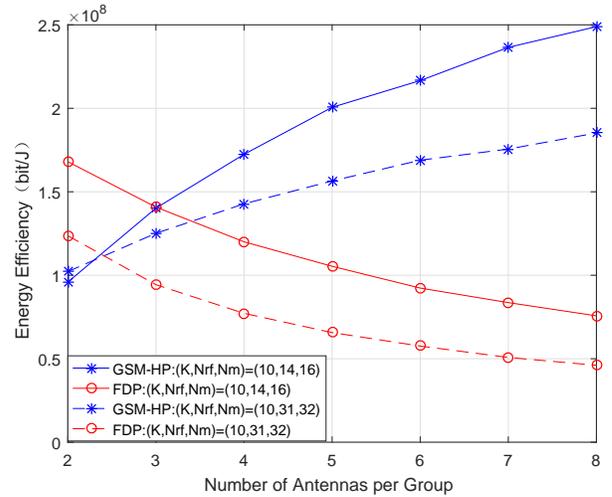}}
\caption{Energy efficiency VS. number of antennas per group of different schemes.}
\label{fig4}
\end{figure}

Fig.~\ref{fig4} depicts the relationship between the EE and the number of antennas per group. Note that FDP does not group the antennas, curves of the FDP scheme show the results of the same number of total antennas as the corresponding GSM scheme. In Fig.~\ref{fig4}, the EE of GSM-HP increases with the increase in the number of antennas per group, while FDP decreases. According to the EE analysis in Section III, more antennas per group only introduces the energy consumption of channel estimation, solution of precoding matrices, and analog phase shifters in GSM-HP. While the times of mm-Wave channel estimation is small, resulting in the energy consumption of channel estimation and solution of precoding matrices being very low. Besides, the energy consumption of analog phase shifters can be ignored compared with the computation power. As a result, the increase in the number of antennas per group does not call for excessive energy consumption, but improves the capacity significantly, which contributes to higher EE.

\begin{figure}[htbp]
\centerline{\includegraphics[width=0.5\textwidth]{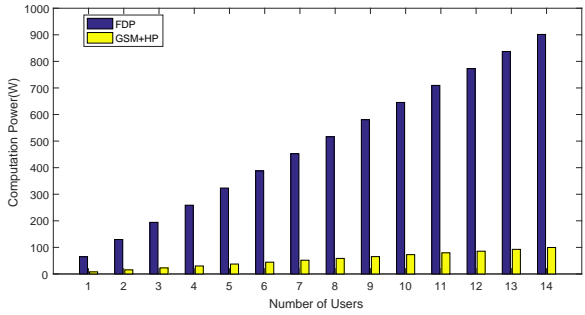}}
\caption{Computation power VS. number of users of different schemes
(${{N}_\text{RF}}$,${{N}_\text{M}}$,${{N}_\text{K}}$)=(14,16,8).}
\label{fig5}
\end{figure}

Furthermore, to have a deeper view of what effect the proposed GSM-HP scheme will have on the computation power, we illustrate computation power versus the number of users in Fig.~\ref{fig5}. As shown in Fig.~\ref{fig5}, the computation power increases sharply  for both schemes along with more users. But the GSM-HP scheme can economize 88\% power consumption compared with the FDP scheme, which shows great promise in improving the EE of 5G cellular networks.

\section{Conclusion}
In this paper, we investigate the EE of the GSM with sub-array HP scheme in the mm-Wave multi-user massive MIMO scene. The EE, considering the computation power, is modeled according to the GSM-HP system. Moreover, the relationships between the EE and the number of users, RF chains and antennas are analyzed. Simulation results show that, combining both GSM and HP technologies can lower the computation power by reducing the number of RF chains. The GSM-HP scheme can improve the EE compared with the traditional full-digital zero-forcing precoding scheme.

\section*{Acknowledgment}
The authors would like to acknowledge the support from National Key R$\&$D Program of China (2016YFE0133000): EU-China study on IoT and 5G (EXICITING-723227)

\end{document}